\documentclass[%
reprint, prl,
showpacs,
 amsmath,amssymb,
]{revtex4-1}
\usepackage{float}
\usepackage{graphicx}
\usepackage{dcolumn}
\usepackage{bm}
\usepackage[utf8]{inputenc}
\usepackage{braket}
\usepackage{hyperref}
\begin{document}

\title{Observation of Local Symmetry in a Photonic System}

\author{Nora Schmitt$^1$}
\author{Steffen Weimann$^1$}
\author{Christian V. Morfonios$^2$}
\author{\\Malte Röntgen$^2$}
\author{Matthias Heinrich$^1$}
\author{Peter Schmelcher$^{2,3}$}
\author{Alexander Szameit$^1$}

 \affiliation{$^1$Institute of Physics, University of Rostock, Albert-Einstein-Str. 23, 18059 Rostock, Germany.}
\affiliation{$^2$Centre for Optical Quantum Technologies}
\affiliation{$^3$Hamburg Centre for Ultrafast Imaging, University of Hamburg, Luruper Chaussee 149, 22761 Hamburg, Germany.}

\date{\today}

\begin{abstract}
	

The concept of local symmetry is a powerful tool in predicting complex transport phenomena in aperiodic media. A nonlocal continuity formalism reveals how local symmetries are encoded into the dynamics of light propagation in discrete waveguide arrays governed by a Schrödinger equation. However, the experimental demonstration is elusive so far. We fabricate representative examples of locally symmetric, globally symmetric and fully non-symmetric configurations in fs laser-written photonic arrays and probe their dynamics. Our approach allows to distinguish all three types of structures.

\end{abstract}


\maketitle

\begin{figure}[t]
	\centering
	\includegraphics[width=\columnwidth]{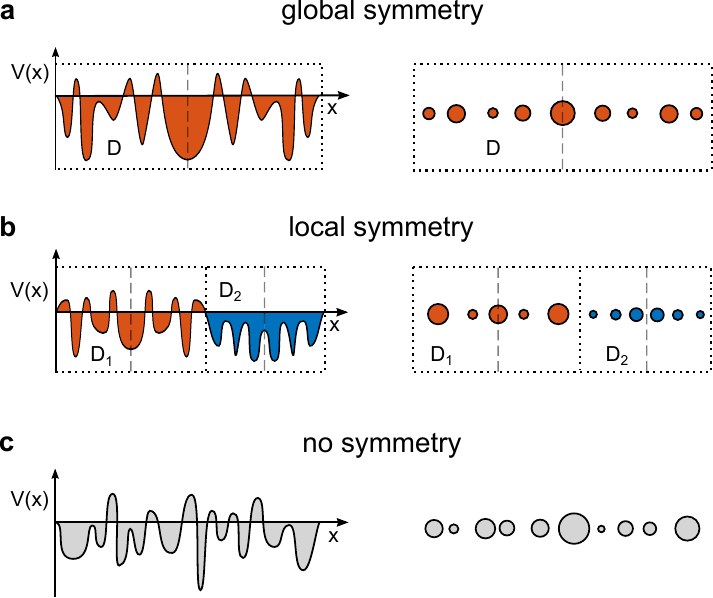}
	\caption{
		Possible gradations of symmetry of a continuous potential (left) and discrete sites (right). The positions of the symmetry axes are indicated by dashed lines.
		(a) Global inversion symmetry with a single overall symmetry domain $D$ (red).
		(b) Locally symmetric system covered by two symmetry domains $D_1$ (red) and $D_2$ (blue).
		(c) Fully non-symmetric configuration.}
	\label{Fig1}
\end{figure}

Whereas the scenario of perfect global symmetries is only valid in an idealized, special class of systems without any imperfection or symmetry breaking and thus notoriously elusive, local symmetries \cite{Kalozoumis2013a} - sometimes also referred to as hidden \cite{Nava2009} or internal \cite{Huang2001} symmetries - abound in nature. These configurations are characterized by internal spatial limitation as illustrated in Fig. \ref{Fig1}. The resulting new class of systems is more general than quasicrystals \cite{Shechtman1984,Levine1984}, which can always be seen as the projection of a periodic lattice in higher dimensions, whereas the only condition for local symmetry is the existence of at least one spatially domain equipped with symmetry \cite{Kalozoumis2013a}. Prominent examples include not only quasicrystals but also macromolecules \cite{Domagala2008,Pascal2001,Echeverria2010} and additionally systems where the global symmetry is broken because of defects or (partial) disorder \cite{Wochner2009,Kalozoumis2014}. Moreover, local symmetries appear in artificial structures, e.g. acoustic waveguides \cite{Kalozoumis2015,Hladky-Hennion2013} or tailored photonic multilayer-systems \cite{Zhukovsky2010,Peng2002}, where they may lead to the occurrence of perfect transmission resonances \cite{Kalozoumis2013}.\\

Invariance with respect to a symmetry transformation is a fundamental concept in physics, which is closely related to the formulation of conservation laws. For continuous transformations, E. Noether stated already in 1918 that to every differentiable symmetry of the action of a physical system there is a corresponding conservation law \cite{Noether1918}. Famous examples are momentum conservation due to the invariance of physical systems with respect to spatial translation or energy conservation due to invariance with respect to time translation.
Symmetry-induced conservation laws of discrete transformations can usually be described by means of the commutation of the corresponding operators with the Hamiltonian. As a consequence, reflection symmetry imposes definite parity and finite translation symmetry imposes Bloch momentum on the eigenstates of the Hamiltonian, characterizing the overall dynamics.

The Hamiltonian of a locally symmetric scenario does in general not commute with the local symmetry operation, even though the potential remains invariant under the respective transformation. Thus, the usual rules of symmetry-induced eigenvalues (such as parity and Bloch momenta) of common eigenstates do not apply. Therefore, tracking the influence of local symmetry on a system's behavior becomes challenging. A promising approach to decode the presence of underlying local symmetries from a system's state is given within a recently developed framework of symmetry-adapted, ``nonlocal currents'' \cite{Kalozoumis2013a, Kalozoumis2014,Kalozoumis2015a,Roentgen2017,Morfonios2017}. In the case of stationary states, these currents are constant within any local symmetry domain and provide an amplitude mapping between symmetry-related points \cite{Kalozoumis2014,Roentgen2017}. This generalizes the usual Bloch and parity theorems to local symmetry.
For a general wave-packet, the nonlocal currents vary in space and time. Nevertheless they have been shown to obey a generalized, local-symmetry-adapted continuity equation \cite{Morfonios2017}, readily represented for discrete models governed by a Schrödinger equation. However, the fact that this approach requires access to the full spatiotemporal information of the complex-valued wave function has thus far prevented any experimental demonstration.\\

\begin{figure}[t]
	\centering
	\includegraphics[width=0.8\columnwidth]{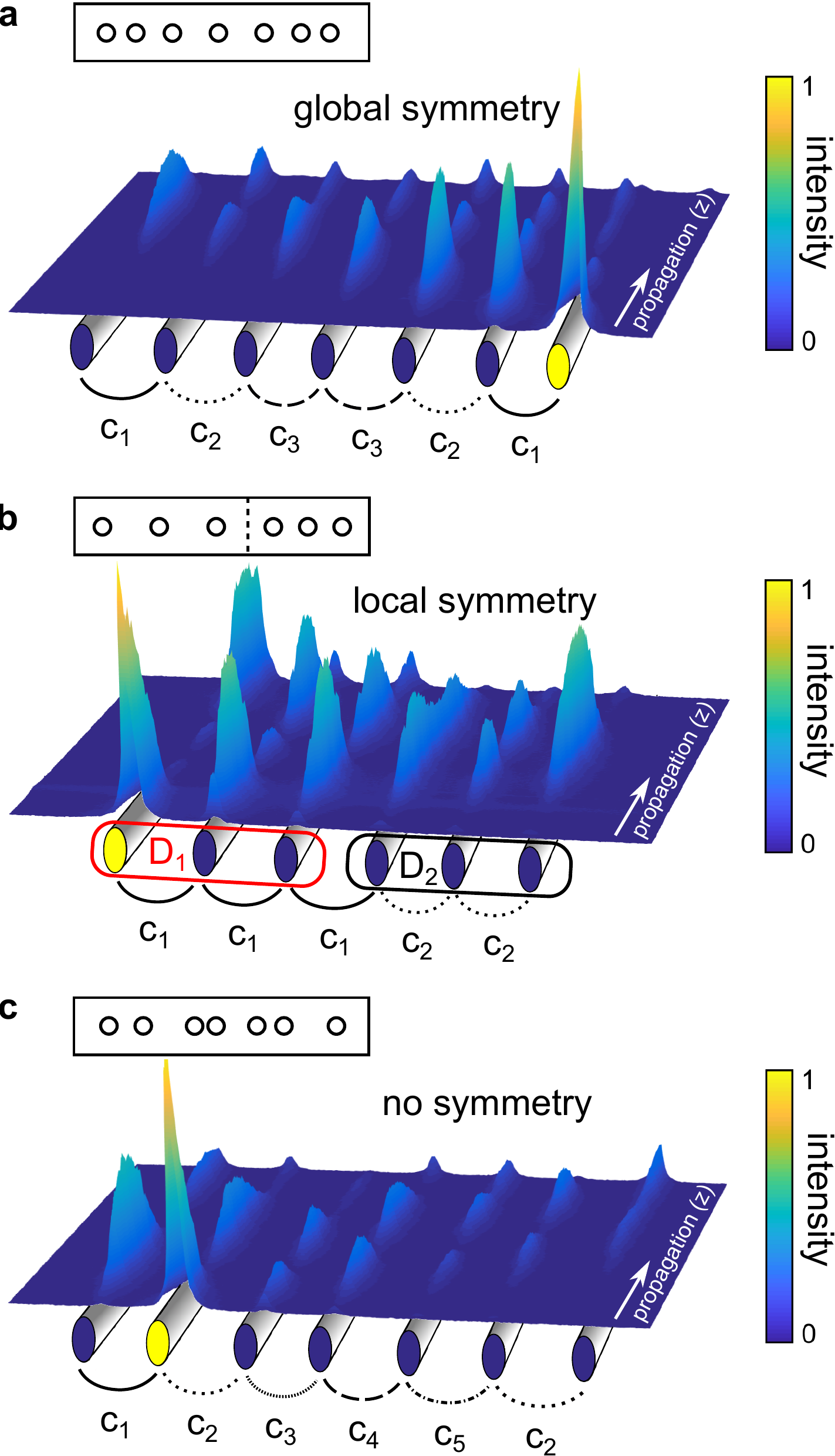}
	\caption{
		Measured fluorescence intensity patterns after single-waveguide excitation.
		(a) Globally symmetric array, probed via single-site excitation of the seventh waveguide (marked yellow).
		The waveguide separations and couplings $c_1$, $c_2$, $c_3$ exhibit inversion symmetry.
		(b) Locally symmetric array, probed via single-site excitation of the first waveguide.
		It can be divided into two domains $D_1$ and $D_2$ with inversion-symmetric configurations of the couplings $c_1$ and $c_2$.
		(c) Non-symmetric array containing five different couplings $c_1$, $c_2$, $c_3$, $c_4$, $c_5$, probed via single site excitation of the second waveguide.
		In each case, the inset on the top left illustrates the lattice geometries with highly exaggerated differences in spacings.
}
	\label{Fig2}
\end{figure}

\begin{figure}[t]
 	\centering
 	\includegraphics[width=\columnwidth]{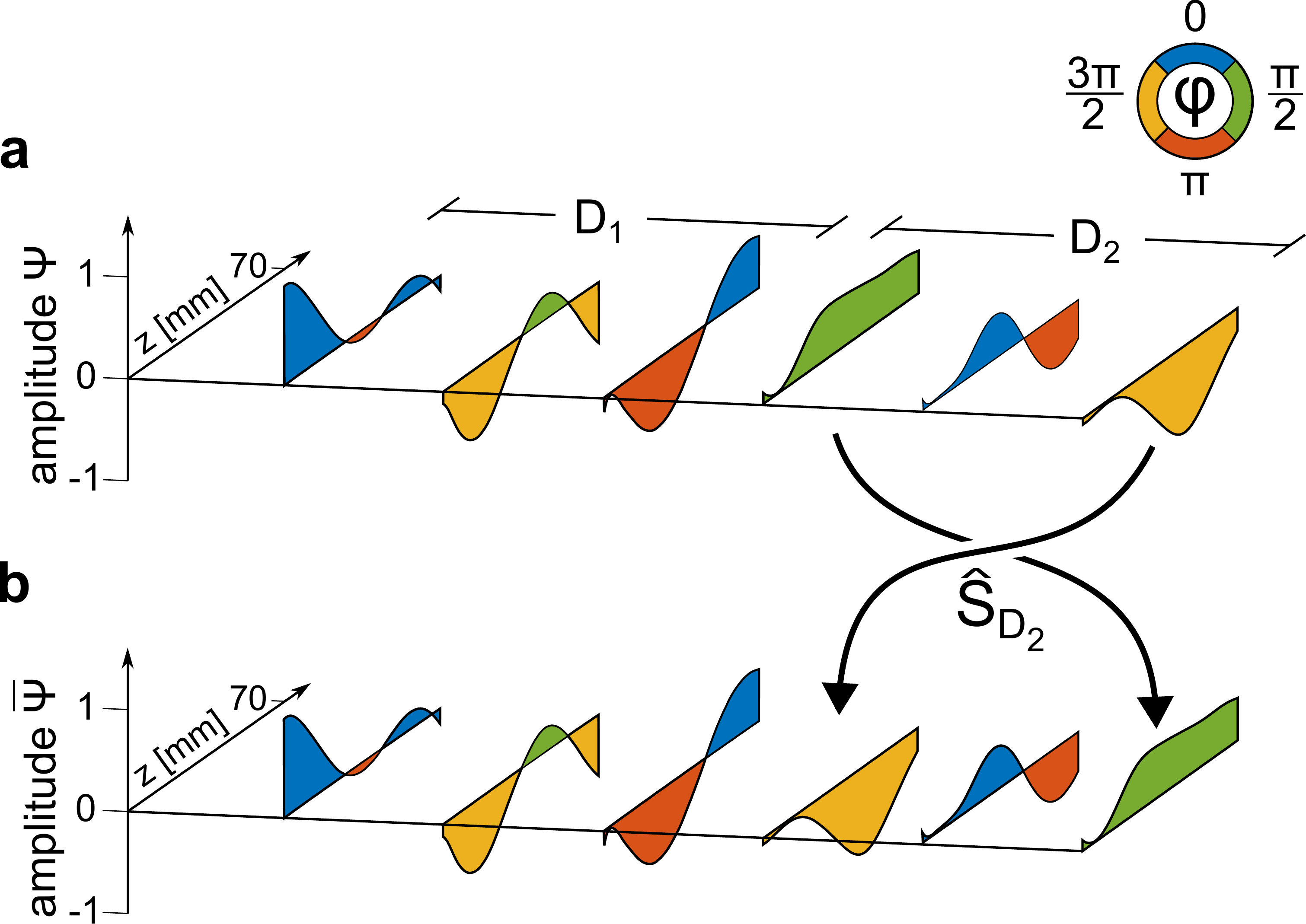}
 	\caption{Relation between wave function $\psi$, local symmetry transformation $\hat S_D$ and transformed wave function $\bar \psi$ of domain $D_2$ of the locally symmetric configuration. (a) Evolution of the fit wave function in the entire system. The phase $\varphi$ is encoded using the given colors. (b) Symmetry transformed wave function $\bar \psi$, obtained by the local symmetry transformation $\hat{S}_{D_2}$ acting on the wave function in domain $D_2$.}
 	\label{Fig3}
 \end{figure}

In this work, we distinguish locally symmetric structures from both fully non-symmetric systems and globally symmetric structures by means of the nonlocal continuity formalism. To address this challenge experimentally we employ the femtosecond laser direct writing technique to fabricate representative examples of locally symmetric, globally symmetric and fully non-symmetric photonic lattices in fused silica glass wafers \cite{Szameit2010}. The symmetries were incorporated in the structures by appropriately tuning the waveguide separation and thus the coupling between adjacent sites in line with the desired distribution (see Fig. \ref{Fig2} - insets top left). The refractive index increase was chosen to be the same for all waveguides to achieve a system with equal on-site potential in order to preserve the phase relation of $\pi/2$ between adjacent waveguides, which is crucial for retrieving the full wave function from intensity-only fluorescence measurements of the light propagation in our waveguide arrays. The corresponding system dynamics was probed via coherent single-site excitation (see Fig. \ref{Fig2}).\\
The complex wave function was extracted from the experimentally observed intensity pattern in accordance with the $\pi/2$ phase jump between adjacent sites and inferring zero crossings at every minimum close to zero. The assumption was justified by tight-binding simulations of the state evolution in our waveguide arrays with the experimentally determined couplings. Thereafter, the wave function was fitted with a high order polynomial in order to allow the calculation of meaningful derivatives. The resulting evolution of the wave function $\psi$ in the locally symmetric system is exemplarily shown in Fig. \ref{Fig3} (a). \\

\begin{figure*}[t]
	\centering
	\includegraphics[width=\textwidth]{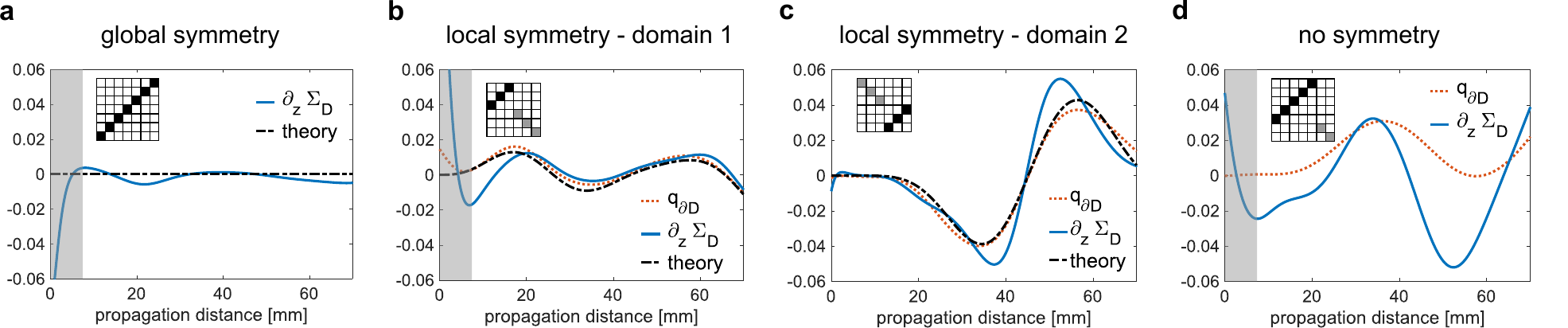}
	\caption{Measurement-based nonlocal continuity equations. (a) Global inversion symmetry. The assumed symmetry transformation is given by an anti-diagonal matrix (top left). As the nonlocal boundary currents are vanishing, the derivative of the nonlocal charge $\partial_z \Sigma_D$ along the propagation (solid blue) is compared to the expected value of zero (dashed black). (b) Local symmetry. $q_{\partial D}$ (dotted red) is nearly equal to $\partial_z \Sigma_D$ (solid blue), as predicted (dashed black) by the continuity equation. The results shown are taken from the measurements in domain $D_1$ of the locally symmetric structure. (c) Same as (b), but for domain $D_2$ of our locally symmetric arrangement. (d) In case of no symmetry, there are no valid symmetry transformations. Any transformation can be chosen and shown to violate the continuity equation because $\partial_z \Sigma_D$ (solid blue) and the nonlocal boundary current $q_{\partial D}$ (dotted red) deviate. }
	\label{Fig4}
\end{figure*}

To reveal the encoding of local symmetries in the state evolution, the discrete local current-density continuity \cite{Boykin2010} for a general state $\ket{\psi}$ is generalized, taking a local symmetry transformation into account \cite{Morfonios2017}. Applying a transformation $\hat{S}_D$ on any quantity - sites or states - is denoted by a bar above the symbol, e.g. $\hat S_D \ket{\psi}=\ket{\bar \psi}$, which is illustrated in Fig. \ref{Fig3} (b). The transformations may be represented by matrices (see insets in Fig. \ref{Fig4}) that interchange the amplitudes of symmetry related sites when they act on a quantity represented by a vector. \\

The local overall discrete probability density $\rho_{D}$ distributed over the sites $n$ in domain $D$ is given by the expectation value of the local density operator $\hat\rho_n=\ket{n}\bra{n}$.
\begin{equation}
\rho_{D}=\sum_{n\in D} \braket{\psi\ket{n}\bra{n}\psi}=\sum_{n\in D} \psi^*_n\psi_{n}=\braket{\psi_D|\psi_D}.
\label{loc_dens}
\end{equation}
In the symmetry-adapted formulation, the local density operator $\hat\rho_n=\ket{n}\bra{n}$ at each site $n$ is replaced by $\ket{n}\bra{\bar n}$. The total nonlocal charge $\Sigma_D$ is calculated by taking the scalar product of the wave function and the symmetry-transformed wave function in the respective domain. \cite{Morfonios2017}
  \begin{equation}
\Sigma_D:=\sum_{n\in D} \braket{\psi\ket{n}\bra{\bar n}\psi}=\sum_{n\in D} \psi^*_n\psi_{\bar n}=\braket{\psi_D|\bar\psi_D}.
\label{nl_ch}
 \end{equation}
 ``Nonlocal'' refers to the influence of non-adjacent but symmetry-related sites on the respective quantity. Note that the only difference between Eq. \ref{loc_dens} and \ref{nl_ch} is given by the symmetry transformation, denoted by a bar above the symbols. Figure \ref{Fig3} illustrates how the local symmetry transformation $\hat S_{D_2}$ acts on the wave function in domain $D_2$ of the locally symmetric system.\\

 The discrete local probability current $j_{n,m}$ between sites $n$ and $m$ is given by \cite{Boykin2010}
  \begin{equation}
j_{n,m}=-i\left(\psi_n^*c_{n,m}\psi_m-\psi_m^*c_{n,m}^*\psi_n\right),
\label{loc_curr}
 \end{equation}
 where $c_{n,m}$ is the coupling between the adjacent sites $n$ and $m=n\pm1$. In the generalized case, the local current $j_{n,m}$ is replaced by the nonlocal (symmetry adapted) current $q_{n,m}$ \cite{Morfonios2017}:
   \begin{equation}
 q_{n,m}=-i\left(\psi_n^*c_{\bar n,\bar m}\psi_{\bar m}-\psi_m^*c_{n,m}^*\psi_{\bar n}\right),
 \label{nl_curr}
 \end{equation}
 Again, the only modification from Eq. \ref{loc_curr} to \ref{nl_curr} are the symmetry transformed sites $\bar n$ and $\bar m$. \\

For arrangements without on-site asymmetry in the refractive index distribution, a simple continuity equation relates the nonlocal charge $\Sigma_D$ to the nonlocal boundary current $q_{\partial D}$, which is ``flowing out of'' each symmetry domain \cite{Morfonios2017} (indices from $a$ to $b$).
\begin{equation}
\partial_z\Sigma_D=q_{a,a-1}+q_{b,b+1}=q_{\partial D},
\end{equation}
where $\partial_z$ is the derivative in propagation direction. \\

The nonlocal boundary current vanishes identically for an even number of sites in one domain of our specific tight-binding Schrödinger system with equal on-site potential and thus we have a well defined phase relation of $\pi/2$ between adjacent sites (see Supplemental material). However, this is a necessary condition to retrieve the full wave function from intensity-only fluorescence measurements. Thus we investigate only domains with an odd number of sites (indices from $a$ to $b$) to allow for a distinction between global and local symmetry. For global and local symmetry, the boundary currents $q_{\partial D}$ are given by (see Supplemental material):
    \begin{equation}
  q_{\partial D} = \pm 2 \left( c_{b,b+1} |\psi_a||\psi_{b+1}| \mp c_{a,a-1} |\psi_b||\psi_{b-1}|\right)
  \label{curr}
  \end{equation}
Due to the fact that the symmetry domain extends over the entire system for all globally inversion symmetric systems, there is inherently no coupling across domain boundaries. Since then $c_{b,b+1} = c_{a,a-1} = 0$ by definition, the nonlocal boundary current vanishes identically for global symmetry. Nevertheless, the continuity equation is still fulfilled, leading to a vanishing derivative of the nonlocal charge.\\

In contrast to the locally and globally symmetric configurations, the fully non-symmetric system by definition entirely lacks symmetry domains. Any arbitrary transformation can be chosen and proven to be no (local) symmetry transformation by showing a violation of the continuity equation. As a result, globally symmetric, locally symmetric and fully non-symmetric configurations may be distinguished by means of the different forms of their corresponding nonlocal continuity equation:
\begin{equation}
\begin{aligned}\begin{split}
\text{global symmetry: }&	\partial_z \Sigma_D=q_{\partial D}=0\\
\text{local symmetry: }&	\partial_z \Sigma_D=q_{\partial D} \\
\text{no symmetry: }&	\partial_z \Sigma_D\neq q_{\partial D}
\end{split}\end{aligned}
\label{vgl}
\end{equation}

To evaluate the continuity equation in the three different experimental configurations, the nonlocal charges, their derivatives and the nonlocal currents were subsequently calculated from the fitted wave function, symmetry transformation and couplings.\\

In case of global symmetry, the symmetry operation may be described by the anti-diagonal matrix shown on the top left of Fig. \ref{Fig4} (a). Because $q_{\partial D}$ is identically zero, it is sufficient to calculate the derivative of the nonlocal charge $\partial_z \Sigma_D$ from the experimental data and compare the result to the expected value of zero. Our experiments and calculations show that the nonlocal continuity equation is fulfilled and indeed vanishing for the globally symmetric configuration (Fig. \ref{Fig4} (a)). In fact, the deviation of $\partial_z\Sigma_D$ from zero for global symmetry may serve as a measure of the validity of our method to retrieve the wave function. Note that the substantial deviations during the first few millimeters of propagation are an artefact of the measurement method. In order to observe the intensity propagation pattern, we employed a fluorescence method that converts a small fraction of the propagating light into omnidirectional light. Although light was injected into the respective waveguide by focussing a laser beam down to an appropriate spot size, the non-unity overlap between focal spot and mode field results in the presence of stray background light that propagates through the sample and may likewise excite fluorescence whenever it traverses a waveguide, and thereby distort the observed pattern. Since this systematic perturbation rapidly dissipates, the subsequent evolution and the $\partial_z\sum_D$ extracted from it coincides remarkably well with the expected behavior. In the further evolution, $\partial_z\Sigma_D$ is remarkably close to the expected value of zero.\\

The locally symmetric system is divided into two inversion symmetric domains -- each containing three sites (see Fig. \ref{Fig2} (b)). In Fig. \ref{Fig4}, the nonlocal current and derivative of the nonlocal charge extracted from the experiment are shown for domain $D_1$ (b) ad $D_2$ (c). Apart from the first few millimeters, the components of the nonlocal continuity equation for the locally symmetric system are in good agreement with the theoretical value as well as with each other. In the second symmetry domain of the locally symmetric system, the extrema of the experimentally determined derivative of the nonlocal charge $\partial_z\Sigma_D$ appear slightly displaced and exaggerated compared to the extracted nonlocal boundary current $q_{\partial D}$ and the theoretical value.
We attribute this feature to the method of flipping the extracted wave function amplitude around zero, which may exaggerate the slope of the zero transitions, as well as to the naturally much higher sensitivity of the derivative to small perturbations. Considering the overall evolution measured, though, our results provide clear evidence that the nonlocal continuity equation is fulfilled in both local symmetry domains.\\

For the fully non-symmetric system, to demonstrate the violation of the nonlocal continuity equation, an applied inversion operation of the first five sites (top left, Fig. \ref{Fig4} (d) is exemplary shown. The experimentally observed evolution of the nonlocal boundary current $q_{\partial D}$ deviates drastically overall from the evolution of the derivative of the nonlocal charge $\partial_z \Sigma_D$. To explicitly prove that the system is fully non-symmetric, one would need show a violation of the continuity equation for all possible transformations. This was done for all possible inversion symmetries with an odd number of sites (see supplemental material).\\

In conclusion, we investigated three arrangements with different gradations of symmetry - global, local and fully non-symmetric - in laser written waveguide arrays, employing the nonlocal continuity approach. We thereby decoded the presence of local symmetries in the Hamiltonian from the generic dynamics of wave packets in a discrete system. We were able to verify the nonlocal continuity equation and to distinguish the three different classes of symmetry by means of their characteristic version of the nonlocal continuity equation.\\

While here demonstrated for the case of inversion symmetry in one dimension, the same formalism can be readily applied and extended to other symmetry transformations and in higher dimensions. The subject offers various possibilities for further experiments, e.g. the engineering of perfect transmission resonances \cite{Kalozoumis2013} or Floquet states in periodically driven setups \cite{Wulf2016,Morfonios2017}, as well as locally symmetric non-Hermitian systems \cite{Kalozoumis2016,Morfonios2017}. Our results constitute the first step in investigating local symmetries in photonic systems and harnessing them to shape the flow of light.\\

The authors gratefully acknowledge financial support from the Deutsche Forschungsgemeinschaft(grants Schm 885/29-1, SZ 276/9-1, SZ 276/12-1, BL 574/13-1, SZ 276/15-1, SZ 276/19-1, and SZ 276/20-1) and the Alfried Krupp von Bohlen und Halbach Foundation. M.\,R. thanks the `Stiftung der deutschen Wirtschaft' for financial support in the framework of a scholarship. The Authors would also like to thank C. Otto for preparing the high-quality fused silica samples used in all experiments presented here.


\begin{thebibliography}{23}%
\makeatletter
\providecommand \@ifxundefined [1]{%
 \@ifx{#1\undefined}
}%
\providecommand \@ifnum [1]{%
 \ifnum #1\expandafter \@firstoftwo
 \else \expandafter \@secondoftwo
 \fi
}%
\providecommand \@ifx [1]{%
 \ifx #1\expandafter \@firstoftwo
 \else \expandafter \@secondoftwo
 \fi
}%
\providecommand \natexlab [1]{#1}%
\providecommand \enquote  [1]{``#1''}%
\providecommand \bibnamefont  [1]{#1}%
\providecommand \bibfnamefont [1]{#1}%
\providecommand \citenamefont [1]{#1}%
\providecommand \href@noop [0]{\@secondoftwo}%
\providecommand \href [0]{\begingroup \@sanitize@url \@href}%
\providecommand \@href[1]{\@@startlink{#1}\@@href}%
\providecommand \@@href[1]{\endgroup#1\@@endlink}%
\providecommand \@sanitize@url [0]{\catcode `\\12\catcode `\$12\catcode
  `\&12\catcode `\#12\catcode `\^12\catcode `\_12\catcode `\%12\relax}%
\providecommand \@@startlink[1]{}%
\providecommand \@@endlink[0]{}%
\providecommand \url  [0]{\begingroup\@sanitize@url \@url }%
\providecommand \@url [1]{\endgroup\@href {#1}{\urlprefix }}%
\providecommand \urlprefix  [0]{URL }%
\providecommand \Eprint [0]{\href }%
\providecommand \doibase [0]{http://dx.doi.org/}%
\providecommand \selectlanguage [0]{\@gobble}%
\providecommand \bibinfo  [0]{\@secondoftwo}%
\providecommand \bibfield  [0]{\@secondoftwo}%
\providecommand \translation [1]{[#1]}%
\providecommand \BibitemOpen [0]{}%
\providecommand \bibitemStop [0]{}%
\providecommand \bibitemNoStop [0]{.\EOS\space}%
\providecommand \EOS [0]{\spacefactor3000\relax}%
\providecommand \BibitemShut  [1]{\csname bibitem#1\endcsname}%
\let\auto@bib@innerbib\@empty
\bibitem [{\citenamefont {Kalozoumis}\ \emph
  {et~al.}(2013{\natexlab{a}})\citenamefont {Kalozoumis}, \citenamefont
  {Morfonios}, \citenamefont {Diakonos},\ and\ \citenamefont
  {Schmelcher}}]{Kalozoumis2013a}%
  \BibitemOpen
  \bibfield  {author} {\bibinfo {author} {\bibfnamefont {P.~A.}\ \bibnamefont
  {Kalozoumis}}, \bibinfo {author} {\bibfnamefont {C.}~\bibnamefont
  {Morfonios}}, \bibinfo {author} {\bibfnamefont {F.~K.}\ \bibnamefont
  {Diakonos}}, \ and\ \bibinfo {author} {\bibfnamefont {P.}~\bibnamefont
  {Schmelcher}},\ }\href@noop {} {\bibfield  {journal} {\bibinfo  {journal}
  {Physical Review A}\ }\textbf {\bibinfo {volume} {87}},\ \bibinfo {pages}
  {032113} (\bibinfo {year} {2013}{\natexlab{a}})}\BibitemShut {NoStop}%
\bibitem [{\citenamefont {Nava}\ \emph {et~al.}(2009)\citenamefont {Nava},
  \citenamefont {Tagüe{\~{n}}a-Mart{\'{\i}}nez}, \citenamefont {del
  R{\'{\i}}o},\ and\ \citenamefont {Naumis}}]{Nava2009}%
  \BibitemOpen
  \bibfield  {author} {\bibinfo {author} {\bibfnamefont {R.}~\bibnamefont
  {Nava}}, \bibinfo {author} {\bibfnamefont {J.}~\bibnamefont
  {Tagüe{\~{n}}a-Mart{\'{\i}}nez}}, \bibinfo {author} {\bibfnamefont
  {J.}~\bibnamefont {del R{\'{\i}}o}}, \ and\ \bibinfo {author} {\bibfnamefont
  {G.}~\bibnamefont {Naumis}},\ }\href@noop {} {\bibfield  {journal} {\bibinfo
  {journal} {Journal of Physics: Condensed Matter}\ }\textbf {\bibinfo {volume}
  {21}},\ \bibinfo {pages} {155901} (\bibinfo {year} {2009})}\BibitemShut
  {NoStop}%
\bibitem [{\citenamefont {Huang}\ \emph {et~al.}(2001)\citenamefont {Huang},
  \citenamefont {Jiang}, \citenamefont {Peng},\ and\ \citenamefont
  {Hu}}]{Huang2001}%
  \BibitemOpen
  \bibfield  {author} {\bibinfo {author} {\bibfnamefont {X.}~\bibnamefont
  {Huang}}, \bibinfo {author} {\bibfnamefont {S.}~\bibnamefont {Jiang}},
  \bibinfo {author} {\bibfnamefont {R.}~\bibnamefont {Peng}}, \ and\ \bibinfo
  {author} {\bibfnamefont {A.}~\bibnamefont {Hu}},\ }\href@noop {} {\bibfield
  {journal} {\bibinfo  {journal} {Physical Review B}\ }\textbf {\bibinfo
  {volume} {63}},\ \bibinfo {pages} {245104} (\bibinfo {year}
  {2001})}\BibitemShut {NoStop}%
\bibitem [{\citenamefont {Shechtman}\ \emph {et~al.}(1984)\citenamefont
  {Shechtman}, \citenamefont {Blech}, \citenamefont {Gratias},\ and\
  \citenamefont {Cahn}}]{Shechtman1984}%
  \BibitemOpen
  \bibfield  {author} {\bibinfo {author} {\bibfnamefont {D.}~\bibnamefont
  {Shechtman}}, \bibinfo {author} {\bibfnamefont {I.}~\bibnamefont {Blech}},
  \bibinfo {author} {\bibfnamefont {D.}~\bibnamefont {Gratias}}, \ and\
  \bibinfo {author} {\bibfnamefont {J.}~\bibnamefont {Cahn}},\ }\href@noop {}
  {\bibfield  {journal} {\bibinfo  {journal} {Physical Review Letters}\
  }\textbf {\bibinfo {volume} {53}},\ \bibinfo {pages} {1951} (\bibinfo {year}
  {1984})}\BibitemShut {NoStop}%
\bibitem [{\citenamefont {Levine}\ and\ \citenamefont
  {Steinhardt}(1984)}]{Levine1984}%
  \BibitemOpen
  \bibfield  {author} {\bibinfo {author} {\bibfnamefont {D.}~\bibnamefont
  {Levine}}\ and\ \bibinfo {author} {\bibfnamefont {P.}~\bibnamefont
  {Steinhardt}},\ }\href@noop {} {\bibfield  {journal} {\bibinfo  {journal}
  {Physical Review Letters}\ }\textbf {\bibinfo {volume} {53}},\ \bibinfo
  {pages} {2477} (\bibinfo {year} {1984})}\BibitemShut {NoStop}%
\bibitem [{\citenamefont {Domaga{\l}a}\ and\ \citenamefont
  {Jelsch}(2008)}]{Domagala2008}%
  \BibitemOpen
  \bibfield  {author} {\bibinfo {author} {\bibfnamefont {S.}~\bibnamefont
  {Domaga{\l}a}}\ and\ \bibinfo {author} {\bibfnamefont {C.}~\bibnamefont
  {Jelsch}},\ }\href@noop {} {\bibfield  {journal} {\bibinfo  {journal}
  {Journal of Applied Crystallography}\ }\textbf {\bibinfo {volume} {41}},\
  \bibinfo {pages} {1140} (\bibinfo {year} {2008})}\BibitemShut {NoStop}%
\bibitem [{\citenamefont {Pascal}(2001)}]{Pascal2001}%
  \BibitemOpen
  \bibfield  {author} {\bibinfo {author} {\bibfnamefont {R.}~\bibnamefont
  {Pascal}},\ }\href@noop {} {\bibfield  {journal} {\bibinfo  {journal} {The
  Journal of Physical Chemistry A}\ }\textbf {\bibinfo {volume} {105}},\
  \bibinfo {pages} {9040} (\bibinfo {year} {2001})}\BibitemShut {NoStop}%
\bibitem [{\citenamefont {Echeverr{\'{\i}}a}\ \emph {et~al.}(2010)\citenamefont
  {Echeverr{\'{\i}}a}, \citenamefont {Carreras}, \citenamefont {Casanova},
  \citenamefont {Alemany},\ and\ \citenamefont {Alvarez}}]{Echeverria2010}%
  \BibitemOpen
  \bibfield  {author} {\bibinfo {author} {\bibfnamefont {J.}~\bibnamefont
  {Echeverr{\'{\i}}a}}, \bibinfo {author} {\bibfnamefont {A.}~\bibnamefont
  {Carreras}}, \bibinfo {author} {\bibfnamefont {D.}~\bibnamefont {Casanova}},
  \bibinfo {author} {\bibfnamefont {P.}~\bibnamefont {Alemany}}, \ and\
  \bibinfo {author} {\bibfnamefont {S.}~\bibnamefont {Alvarez}},\ }\href@noop
  {} {\bibfield  {journal} {\bibinfo  {journal} {Chemistry - A European
  Journal}\ }\textbf {\bibinfo {volume} {17}},\ \bibinfo {pages} {359}
  (\bibinfo {year} {2010})}\BibitemShut {NoStop}%
\bibitem [{\citenamefont {Wochner}\ \emph {et~al.}(2009)\citenamefont
  {Wochner}, \citenamefont {Gutt}, \citenamefont {Autenrieth}, \citenamefont
  {Demmer}, \citenamefont {Bugaev}, \citenamefont {Ortiz}, \citenamefont
  {Duri}, \citenamefont {Zontone}, \citenamefont {Grübel},\ and\ \citenamefont
  {Dosch}}]{Wochner2009}%
  \BibitemOpen
  \bibfield  {author} {\bibinfo {author} {\bibfnamefont {P.}~\bibnamefont
  {Wochner}}, \bibinfo {author} {\bibfnamefont {C.}~\bibnamefont {Gutt}},
  \bibinfo {author} {\bibfnamefont {T.}~\bibnamefont {Autenrieth}}, \bibinfo
  {author} {\bibfnamefont {T.}~\bibnamefont {Demmer}}, \bibinfo {author}
  {\bibfnamefont {V.}~\bibnamefont {Bugaev}}, \bibinfo {author} {\bibfnamefont
  {A.}~\bibnamefont {Ortiz}}, \bibinfo {author} {\bibfnamefont
  {A.}~\bibnamefont {Duri}}, \bibinfo {author} {\bibfnamefont {F.}~\bibnamefont
  {Zontone}}, \bibinfo {author} {\bibfnamefont {G.}~\bibnamefont {Grübel}}, \
  and\ \bibinfo {author} {\bibfnamefont {H.}~\bibnamefont {Dosch}},\
  }\href@noop {} {\bibfield  {journal} {\bibinfo  {journal} {Proceedings of the
  National Academy of Sciences}\ }\textbf {\bibinfo {volume} {106}},\ \bibinfo
  {pages} {11511} (\bibinfo {year} {2009})}\BibitemShut {NoStop}%
\bibitem [{\citenamefont {Kalozoumis}\ \emph {et~al.}(2014)\citenamefont
  {Kalozoumis}, \citenamefont {Morfonios}, \citenamefont {Diakonos},\ and\
  \citenamefont {Schmelcher}}]{Kalozoumis2014}%
  \BibitemOpen
  \bibfield  {author} {\bibinfo {author} {\bibfnamefont {P.~A.}\ \bibnamefont
  {Kalozoumis}}, \bibinfo {author} {\bibfnamefont {C.}~\bibnamefont
  {Morfonios}}, \bibinfo {author} {\bibfnamefont {F.~K.}\ \bibnamefont
  {Diakonos}}, \ and\ \bibinfo {author} {\bibfnamefont {P.}~\bibnamefont
  {Schmelcher}},\ }\href@noop {} {\bibfield  {journal} {\bibinfo  {journal}
  {Physical Review Letters}\ }\textbf {\bibinfo {volume} {113}},\ \bibinfo
  {pages} {050403} (\bibinfo {year} {2014})}\BibitemShut {NoStop}%
\bibitem [{\citenamefont {Kalozoumis}\ \emph
  {et~al.}(2015{\natexlab{a}})\citenamefont {Kalozoumis}, \citenamefont
  {Richoux}, \citenamefont {Diakonos}, \citenamefont {Theocharis},\ and\
  \citenamefont {Schmelcher}}]{Kalozoumis2015}%
  \BibitemOpen
  \bibfield  {author} {\bibinfo {author} {\bibfnamefont {P.~A.}\ \bibnamefont
  {Kalozoumis}}, \bibinfo {author} {\bibfnamefont {O.}~\bibnamefont {Richoux}},
  \bibinfo {author} {\bibfnamefont {F.~K.}\ \bibnamefont {Diakonos}}, \bibinfo
  {author} {\bibfnamefont {G.}~\bibnamefont {Theocharis}}, \ and\ \bibinfo
  {author} {\bibfnamefont {P.}~\bibnamefont {Schmelcher}},\ }\href@noop {}
  {\bibfield  {journal} {\bibinfo  {journal} {Physical Review B}\ }\textbf
  {\bibinfo {volume} {92}},\ \bibinfo {pages} {014303} (\bibinfo {year}
  {2015}{\natexlab{a}})}\BibitemShut {NoStop}%
\bibitem [{\citenamefont {Hladky-Hennion}\ \emph {et~al.}(2013)\citenamefont
  {Hladky-Hennion}, \citenamefont {Vasseur}, \citenamefont {Degraeve},
  \citenamefont {Granger},\ and\ \citenamefont
  {de~Billy}}]{Hladky-Hennion2013}%
  \BibitemOpen
  \bibfield  {author} {\bibinfo {author} {\bibfnamefont {A.~C.}\ \bibnamefont
  {Hladky-Hennion}}, \bibinfo {author} {\bibfnamefont {J.~O.}\ \bibnamefont
  {Vasseur}}, \bibinfo {author} {\bibfnamefont {S.}~\bibnamefont {Degraeve}},
  \bibinfo {author} {\bibfnamefont {C.}~\bibnamefont {Granger}}, \ and\
  \bibinfo {author} {\bibfnamefont {M.}~\bibnamefont {de~Billy}},\ }\href@noop
  {} {\bibfield  {journal} {\bibinfo  {journal} {Journal of Applied Physics}\
  }\textbf {\bibinfo {volume} {113}},\ \bibinfo {pages} {154901} (\bibinfo
  {year} {2013})}\BibitemShut {NoStop}%
\bibitem [{\citenamefont {Zhukovsky}(2010)}]{Zhukovsky2010}%
  \BibitemOpen
  \bibfield  {author} {\bibinfo {author} {\bibfnamefont {S.}~\bibnamefont
  {Zhukovsky}},\ }\href@noop {} {\bibfield  {journal} {\bibinfo  {journal}
  {Physical Review A}\ }\textbf {\bibinfo {volume} {81}},\ \bibinfo {pages}
  {053808} (\bibinfo {year} {2010})}\BibitemShut {NoStop}%
\bibitem [{\citenamefont {Peng}\ \emph {et~al.}(2002)\citenamefont {Peng},
  \citenamefont {Huang}, \citenamefont {Qiu}, \citenamefont {Wang},
  \citenamefont {Hu}, \citenamefont {Jiang},\ and\ \citenamefont
  {Mazzer}}]{Peng2002}%
  \BibitemOpen
  \bibfield  {author} {\bibinfo {author} {\bibfnamefont {R.}~\bibnamefont
  {Peng}}, \bibinfo {author} {\bibfnamefont {X.}~\bibnamefont {Huang}},
  \bibinfo {author} {\bibfnamefont {F.}~\bibnamefont {Qiu}}, \bibinfo {author}
  {\bibfnamefont {M.}~\bibnamefont {Wang}}, \bibinfo {author} {\bibfnamefont
  {A.}~\bibnamefont {Hu}}, \bibinfo {author} {\bibfnamefont {S.}~\bibnamefont
  {Jiang}}, \ and\ \bibinfo {author} {\bibfnamefont {M.}~\bibnamefont
  {Mazzer}},\ }\href@noop {} {\bibfield  {journal} {\bibinfo  {journal}
  {Applied Physics Letters}\ }\textbf {\bibinfo {volume} {80}},\ \bibinfo
  {pages} {3063} (\bibinfo {year} {2002})}\BibitemShut {NoStop}%
\bibitem [{\citenamefont {Kalozoumis}\ \emph
  {et~al.}(2013{\natexlab{b}})\citenamefont {Kalozoumis}, \citenamefont
  {Morfonios}, \citenamefont {Palaiodimopoulos}, \citenamefont {Diakonos},\
  and\ \citenamefont {Schmelcher}}]{Kalozoumis2013}%
  \BibitemOpen
  \bibfield  {author} {\bibinfo {author} {\bibfnamefont {P.~A.}\ \bibnamefont
  {Kalozoumis}}, \bibinfo {author} {\bibfnamefont {C.}~\bibnamefont
  {Morfonios}}, \bibinfo {author} {\bibfnamefont {N.}~\bibnamefont
  {Palaiodimopoulos}}, \bibinfo {author} {\bibfnamefont {F.~K.}\ \bibnamefont
  {Diakonos}}, \ and\ \bibinfo {author} {\bibfnamefont {P.}~\bibnamefont
  {Schmelcher}},\ }\href@noop {} {\bibfield  {journal} {\bibinfo  {journal}
  {Physical Review A}\ }\textbf {\bibinfo {volume} {88}},\ \bibinfo {pages}
  {033857} (\bibinfo {year} {2013}{\natexlab{b}})}\BibitemShut {NoStop}%
\bibitem [{\citenamefont {Noether}(1918)}]{Noether1918}%
  \BibitemOpen
  \bibfield  {author} {\bibinfo {author} {\bibfnamefont {E.}~\bibnamefont
  {Noether}},\ }\href@noop {} {\bibfield  {journal} {\bibinfo  {journal}
  {Nachr. D. König. Gesellsch. D. Wiss. Zu Göttingen}\ } (\bibinfo {year}
  {1918})}\BibitemShut {NoStop}%
\bibitem [{\citenamefont {Kalozoumis}\ \emph
  {et~al.}(2015{\natexlab{b}})\citenamefont {Kalozoumis}, \citenamefont
  {Morfonios}, \citenamefont {Diakonos},\ and\ \citenamefont
  {Schmelcher}}]{Kalozoumis2015a}%
  \BibitemOpen
  \bibfield  {author} {\bibinfo {author} {\bibfnamefont {P.~A.}\ \bibnamefont
  {Kalozoumis}}, \bibinfo {author} {\bibfnamefont {C.~V.}\ \bibnamefont
  {Morfonios}}, \bibinfo {author} {\bibfnamefont {F.~K.}\ \bibnamefont
  {Diakonos}}, \ and\ \bibinfo {author} {\bibfnamefont {P.}~\bibnamefont
  {Schmelcher}},\ }\href@noop {} {\bibfield  {journal} {\bibinfo  {journal}
  {Annals of Physics}\ }\textbf {\bibinfo {volume} {362}},\ \bibinfo {pages}
  {684} (\bibinfo {year} {2015}{\natexlab{b}})}\BibitemShut {NoStop}%
\bibitem [{\citenamefont {Röntgen}\ \emph {et~al.}(2017)\citenamefont
  {Röntgen}, \citenamefont {Morfonios}, \citenamefont {Diakonos},\ and\
  \citenamefont {Schmelcher}}]{Roentgen2017}%
  \BibitemOpen
  \bibfield  {author} {\bibinfo {author} {\bibfnamefont {M.}~\bibnamefont
  {Röntgen}}, \bibinfo {author} {\bibfnamefont {C.~V.}\ \bibnamefont
  {Morfonios}}, \bibinfo {author} {\bibfnamefont {F.~K.}\ \bibnamefont
  {Diakonos}}, \ and\ \bibinfo {author} {\bibfnamefont {P.}~\bibnamefont
  {Schmelcher}},\ }\href@noop {} {\bibfield  {journal} {\bibinfo  {journal}
  {Annals of Physics}\ }\textbf {\bibinfo {volume} {380}},\ \bibinfo {pages}
  {135} (\bibinfo {year} {2017})}\BibitemShut {NoStop}%
\bibitem [{\citenamefont {Morfonios}\ \emph {et~al.}(2017)\citenamefont
  {Morfonios}, \citenamefont {Kalozoumis}, \citenamefont {Diakonos},\ and\
  \citenamefont {Schmelcher}}]{Morfonios2017}%
  \BibitemOpen
  \bibfield  {author} {\bibinfo {author} {\bibfnamefont {C.~V.}\ \bibnamefont
  {Morfonios}}, \bibinfo {author} {\bibfnamefont {P.~A.}\ \bibnamefont
  {Kalozoumis}}, \bibinfo {author} {\bibfnamefont {F.~K.}\ \bibnamefont
  {Diakonos}}, \ and\ \bibinfo {author} {\bibfnamefont {P.}~\bibnamefont
  {Schmelcher}},\ }\href@noop {} {\bibfield  {journal} {\bibinfo  {journal}
  {Annals of Physics}\ }\textbf {\bibinfo {volume} {385}},\ \bibinfo {pages}
  {623} (\bibinfo {year} {2017})}\BibitemShut {NoStop}%
\bibitem [{\citenamefont {Szameit}\ and\ \citenamefont
  {Nolte}(2010)}]{Szameit2010}%
  \BibitemOpen
  \bibfield  {author} {\bibinfo {author} {\bibfnamefont {A.}~\bibnamefont
  {Szameit}}\ and\ \bibinfo {author} {\bibfnamefont {S.}~\bibnamefont
  {Nolte}},\ }\href@noop {} {\bibfield  {journal} {\bibinfo  {journal} {Journal
  of Physics B: Atomic, Molecular and Optical Physics}\ }\textbf {\bibinfo
  {volume} {43}},\ \bibinfo {pages} {163001} (\bibinfo {year}
  {2010})}\BibitemShut {NoStop}%
\bibitem [{\citenamefont {Boykin}\ \emph {et~al.}(2010)\citenamefont {Boykin},
  \citenamefont {Luisier},\ and\ \citenamefont {Klimeck}}]{Boykin2010}%
  \BibitemOpen
  \bibfield  {author} {\bibinfo {author} {\bibfnamefont {T.~B.}\ \bibnamefont
  {Boykin}}, \bibinfo {author} {\bibfnamefont {M.}~\bibnamefont {Luisier}}, \
  and\ \bibinfo {author} {\bibfnamefont {G.}~\bibnamefont {Klimeck}},\
  }\href@noop {} {\bibfield  {journal} {\bibinfo  {journal} {European Journal
  of Physics}\ }\textbf {\bibinfo {volume} {31}},\ \bibinfo {pages} {1077}
  (\bibinfo {year} {2010})}\BibitemShut {NoStop}%
\bibitem [{\citenamefont {Wulf}\ \emph {et~al.}(2016)\citenamefont {Wulf},
  \citenamefont {Morfonios}, \citenamefont {Diakonos},\ and\ \citenamefont
  {Schmelcher}}]{Wulf2016}%
  \BibitemOpen
  \bibfield  {author} {\bibinfo {author} {\bibfnamefont {T.}~\bibnamefont
  {Wulf}}, \bibinfo {author} {\bibfnamefont {C.~V.}\ \bibnamefont {Morfonios}},
  \bibinfo {author} {\bibfnamefont {F.~K.}\ \bibnamefont {Diakonos}}, \ and\
  \bibinfo {author} {\bibfnamefont {P.}~\bibnamefont {Schmelcher}},\
  }\href@noop {} {\bibfield  {journal} {\bibinfo  {journal} {Physical Review
  E}\ }\textbf {\bibinfo {volume} {93}},\ \bibinfo {pages} {052215} (\bibinfo
  {year} {2016})}\BibitemShut {NoStop}%
\bibitem [{\citenamefont {Kalozoumis}\ \emph {et~al.}(2016)\citenamefont
  {Kalozoumis}, \citenamefont {Morfonios}, \citenamefont {Diakonos},\ and\
  \citenamefont {Schmelcher}}]{Kalozoumis2016}%
  \BibitemOpen
  \bibfield  {author} {\bibinfo {author} {\bibfnamefont {P.~A.}\ \bibnamefont
  {Kalozoumis}}, \bibinfo {author} {\bibfnamefont {C.~V.}\ \bibnamefont
  {Morfonios}}, \bibinfo {author} {\bibfnamefont {F.~K.}\ \bibnamefont
  {Diakonos}}, \ and\ \bibinfo {author} {\bibfnamefont {P.}~\bibnamefont
  {Schmelcher}},\ }\href@noop {} {\bibfield  {journal} {\bibinfo  {journal}
  {Physical Review A}\ }\textbf {\bibinfo {volume} {93}},\ \bibinfo {pages}
  {063831} (\bibinfo {year} {2016})}\BibitemShut {NoStop}%
\end{thebibliography}
\end{document}